\documentclass[letterpaper,twocolumn,10pt]{article}
\usepackage{usenix2020_SOUPS}
\usepackage{graphicx}
\usepackage{comment}
\usepackage{tikz}
\usepackage{amsmath}

\usepackage{filecontents}

\begin{document}

\date{}

\title{\Large \bf Towards Inclusive Design for Privacy and Security\\
  Perspectives from an Aging Society}

\def\plainauthor{Pakianathan and Perrault}

\author{
{\rm Pavithren V S Pakianathan}\\
Singapore University of Technology and Design
\and
{\rm Simon Perrault}\\
Singapore University of Technology and Design
} 

\maketitle
\thecopyright

\begin{abstract}
Over the past few years, older adults  in Singapore have been massively connecting to the  Internet using Smartphone.
However due to the ever-changing nature of Technology and Cybersecurity landscape, an older adult's limited technical and Privacy and Security (P \& S) knowledge, experience and declining cognitive and physical abilities puts them at higher risks.
Furthermore mainstream smartphone applications, which are generally not designed with older adults in mind, could result in mismatched mental models thereby creating usability issues. 
We interviewed 10 older adults above 65 and 10 adults assisting them based in Singapore to investigate how smartphone P \& S can be redesigned inclusively by addressing the needs of older adults and people who support them. Our results show that socio-cultural factors affected the process of getting or providing P \& S help, culture and attitude affected learning behaviours and older adults expressed heterogeneous P \& S preferences based on contextual factors and level of convenience, however there are opportunities for the mechanisms to be senior-friendly. Due to the complex relationship between an older adult's milieu and technology, we aim to utilize a technology probe to investigate further and contribute towards an inclusive P \& S model. 
\end{abstract}

\section{Introduction}
The population is aging rapidly around the world and especially in Asia. Half of Singapore's population is expected to be above 50 by the year 2050~\cite{united2017world}. 
Over the last few years, older adults above 60 have been the highest contributors to smartphone adoption in Singapore\cite{IMDA2018}, mainly for social media, media consumption and leisure.
As they “age in place”, while living alone or with family members, smartphones allow them to keep connected with family, friends or keep them engaged.
In most cases, seniors upgraded from a feature phone to a smartphone and have not been properly trained on how to navigate the Internet safely.
The constant emergence of new technologies, applications and P \& S threats ---often with varying levels of complexity--- means that older adults who generally have inadequate experience or have declining physical or cognitive abilities are unable to grapple with these challenges alone. 
Oftentimes, they rely on 'supporters'~\cite{Mecke2018}, who might be tech-savvy family members, friends, shopkeepers, or even strangers to seek for assistance with usability issues with their smartphones.
While prior work has shown that social-help\cite{Wan2019,Aljallad2019,Mendel2019social} is one of the most promising ways to address the P \& S needs of older adults, little has been done to improve existing interfaces to make P \& S threats more salient and educate older adults on cyber-security threats.

We conducted semi-structured interviews with 10 older adults above 65 years old and 10 adults above 18 years old who support older adults with using smartphones.
By including younger stakeholders who ’support’ older adults into our study and specifically looking at P \& S surrounding smartphones, we discovered challenges faced by older adults and their supporters which could be turned into opportunities that could be of interest to smartphone and application developers, designers and policymakers. 

Our work contributes by adding holisticity to previous works by Frik et al.~\cite{Frik2019} and Mecke et al.~\cite{Mecke2018} by investigating the P \& S challenges faced by older adults and their social circle in terms of the smartphone context.
Based on these findings, we propose preliminary design considerations for designing inclusive digital interventions which allow older smartphone users to maintain their P \& S online and at the same time give them a sense of agency and autonomy.
We expect the inclusive design approach to benefit other groups who might not necessarily be of older age.
\section{Related Work}
\subsection{Singapore Context}
Smartphone have become a major part of an older adult's lives in terms of providing a bridge to connect with their loved ones or as a source of entertainment.
Furthermore, phones offer lesser barriers for access to the internet as compared to devices such as a PC and with the ongoing COVID-19 pandemic and "circuit-breaker" safe-distancing measures, seniors are encouraged to remain home and engage in social activities via the Internet\cite{IMDA2020NationalEfforts}.
According to the Infocomm Media Development Authority(IMDA) 2018 Annual Survey On Infocomm Usage In Households And By Individuals~\cite{IMDA2018}, the main uses of smartphones by older adults in Singapore are for instant messaging, social network, general web browsing and downloading or watching movies, or telephoning over VoIP.
However Singaporean older adults exhibit the lowest level of confidence in making online transactions and trust that their privacy is protected while performing online transactions. 
In order to address their issues, the Singapore government has been taking initiatives such as the Digital Clinic Initiative~\cite{IMDA2020} which allows seniors to get one-to-one help from volunteers with regards to getting smartphone related help. While these are encouraging first steps in terms of prevention, more needs to be done as we argue in our paper. 

\subsection{Amplified P \& S issues among Older Adults}

As smartphones get more high tech and feature packed,  applications have gained capability of collecting sensitive information from various sources such as GPS, microphones, camera, SMS, keyboard, browsing history etc. in order to provide functionalities to the user as well as improve their services and serve targeted advertisements. Kelley et al.~\cite{Kelley2012} found that Android users do not understand app permissions and are not aware about the risks of accepting app permissions. 
Frik et al.~\cite{Frik2019} reported that older adults tend to have a reduced P \& S knowledge, as well as misconception about data flows and mitigation strategies which may amplify certain P \& S risks due to barriers to learning and understanding due decline in abilities. 
Others\cite{Chouhan2019,Wan2019,Aljallad2019} have suggested collective intelligence via social help to be a form of support for older adults to resolve the issues they face. 

\subsection{Socio-cultural Factors in P \& S}
As technology permeates an individual's milieu, it has become crucial to evaluate the relationship betwen socio-cultural factors, P \& S and technology. While working on a privacy preserving model in a smart home setting, Kraemer \cite{kraemer2018preserving} proposes looking at privacy from a socio-cultural perspective due to complexities in a smart-home setting and its sensitivity to social factors\cite{nissenbaum2009privacy}. Some researchers have taken advantage of the social factors to develop better P \& S. For example, Chouhan et al.~\cite{Chouhan2019} developed a novel model, Co-oPS that utilizes collective efficacy which relies on community members to crowdsource information and make better P \& S decisions together, however it comes with its own set of challenges and opportunities. Mendel et al.~\cite{Mendel2019mom} brings it a step further by identifying that a closely knit social circle ---a family--- is likely to provide help regarding privacy and security compared to other bigger social groups and have an advantageous position due to their familiarity of an older adult's preferences. Mecke et al.~\cite{Mecke2018} had proposed a framework where older adults can outsource their security to a ‘supporter’ through push or pull arrangement either implicitly or explicitly. Although these are valuable suggestions and developments, due to variances in different cultures around the world, there might be differences in how social support could play out~\cite{Karasawa2011,Wan2019}. 
\section{Interview}
\subsection{Procedure}
The study procedures went through ethics review and were approved by SUTD’s Institutional Review Board(IRB). Participants had to sign informed consent and prior to the study and audio recordings were captured for transcription purposes.

We conducted 1 hour semi-structured interviews with older adults above 65 years old and adults above 18 years old who have assisted them with the use of smartphones.
During the study, we discussed smartphone related P \& S issues, behavioral strategies and the process of learning and teaching about smartphone P \& S between older adults and adults. 

This study was focused on healthy participants residing in Singapore who own a smartphone and know how to read and write English. Audio was recorded with participants' permission for transcription purposes. Participants were provided a S\$10 voucher as reimbursement for their participation.

\begin{figure}[htp]
    \centering
    \includegraphics[width=\columnwidth]{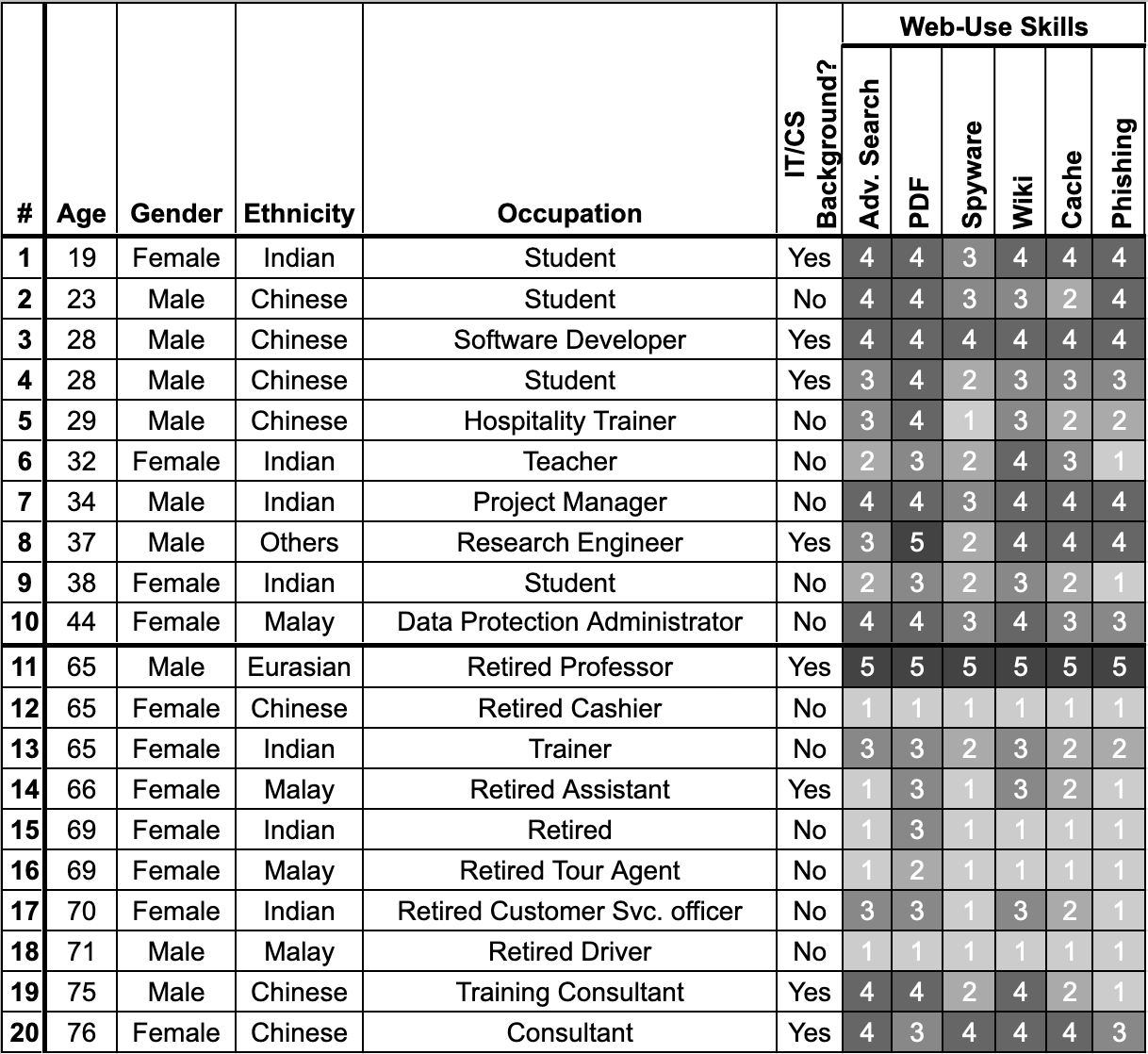}
    \caption{Summary of Participants}
    \label{fig:summary}
\end{figure}

The structure of our interviews was inspired by Frik et al.~\cite{Frik2019} and Zeng et al.~\cite{Zeng2017} with some changes to focus on smartphone usage with our two groups of stakeholders. The semi-structured interview questionnaire was adapted from\cite{Frik2019} to target smartphone usage behaviours and preferences among older adult and adults who received and provided assistance respectively. There were 2 surveys conducted via Google Forms before and after the interview to collect information about the demographic information and technical background by rating their Internet use skills\cite{hargittai2012succinct} based on a scale of 1 (No Understanding) to 5 (Full Understanding). The participants' profile is presented below~(see Figure~\ref{fig:summary}).

\subsection{Analysis}
Themes were generated deductively using thematic analysis from interview transcriptions from voice recordings and data was coded by a single researcher. \cite{nowell2017thematic}. The data collected from the pre and post interview surveys were used to supplement the findings from the semi-structured interviews.
\section{Findings and Discussion}
We present a focused summary of our main findings and the implication they have on P \& S for smartphone applications. Most of our findings corroborated with those by Alisa et. al ~\cite{Frik2019} - P \& S concerns, misconceptions about data practices and learning and mitigation strategies and P \& S trade-offs.
However, by including younger stakeholders who 'support' older adults in our study and specifically looking at the smartphone context, we were able to reveal additional facets to challenges around an older adult's P \& S related issues. Firstly, we identified that older adults were primarily reliant their social circle to get P \& S related information and support. However, the presence of multiple 'supporters' who might not have a complete picture of the older adult's P \& S adds further complexities to the situation. Secondly, attitude and culture were found to influence an older adult's willingness to learn about P \& S. Lastly, they expressed contextualized needs when it comes to making decisions related to Privacy and there emerges a need for older adult-friendly safety mechanisms. Note that quotes may include Singlish(Colloquial Singaporean English) expressions and may come with annotations (using brackets) to clarify. 

\subsection {Social Support for P \& S}
\textbf{Reliance on Family/Friends/Relatives for P \& S support}
Most older adults were primarily reliant on their social circle consisting of family members, relatives or friends to receive help and seek information related to P \& S. (\emph{“For i[nternet] banking, I am a bit nervous lah, a bit worried, in case I touch anything wrong or what ah, so for online banking I also get help from my son.”} - P7, 66 y.o.). (\emph{“.. because of my connections...we have a WhatsApp chat group, they will just post it[P \& S updates] ”} - P18, 75 y.o.).
Although, this might seem positive, their help might not necessary be beneficial as there were few adult participants who recommended older adults unsafe practices such as setting easy to remember passwords(\emph{"I use something that is very close to them[parents], like their IC [Identification Card] number or their phone number, then add on something behind"} - P10, 28 y.o.)., skipping tutorials, not explaining about privacy policies and application permissions. One adult participant even shared that he saves his passwords in his contact lists (\emph{"if someone access my contact lists, they will know all my passwords(giggles)"} - P12, 29 y.o.). This meant that younger 'supporters' could also potentially share their risky behaviours while helping older adults.

\textbf{Family members' opacity about  Older Adults' P \& S}
As older adults rely on various family members for help, it becomes a challenge to track the activities performed by other 'supporters'. (\emph{“I am not sure who installs the apps for grandma it's either me or my cousins, parents, other cousins, or other kids. Depends on who visits her house." }- P16, 23 y.o.). Some of them also shared that the younger family members might not always be available to assist them due to their busy lifestyles, resulting in self-exploration or asking other relatives and friends for help. (\emph{“I have a daughter who is a IT specialist ... but I don't trouble her too much because she's very busy”} - P19, 76 y.o.).
Furthermore, although adults share tips to their parents on safekeeping themselves online, they are not really sure if these advice are fully adhered to completely(\emph{“I think my mom won’t share any private information online.”} - P10, 28 y.o.).  The challenges while relying on multiple stakeholders to receive support and the supporters' assumptions about an older adult's perception of P \& S, reveals several fault lines, and highlight the need for improving systems to give older adults a greater sense of autonomy.

\subsection{Learning about Privacy \& Security}

\textbf{Attitudinal Factors}
Most of the participants from both groups were generally not proactive in learning or updating themselves about P \& S (\emph{“unless I am in security field, I wouldn't go about proactively searching about all these threats.” }- P10, 28 y.o.). One participant(P14, 69 y.o) stated that she was not bothered to keep herself updated about cyber-security related threats. Overall, there seems to be a lackadaisical attitude towards smartphone P \& S such that one of the participant suggested a drastic measure to tackle this(\emph{“Let them get hit once and then they learn their lesson."} - P11, 28 y.o.). 
As the an older adult's main source of information about P \& S was family members or relatives, this meant that they are limited to the knowledge imparted by them and thus may not necessarily receive correct and sufficient information to keep themselves safe online. This emphasizes the needs highlighted by Frik et al.~\cite{Frik2019} to push P \& S related information targeted towards older adults, (\emph{“Simple, visual, bite sized information which are shared via WhatsApp would be useful."}- P8, 44 y.o.). 

\textbf{Cultural Factors}
A common theme arose on why individuals did not want to seek help when they experience issues. Some participants felt that asking for help could result in embarrassment or could trouble their busy children. (\emph{"Some elderly participants [at the Digital Clinic] don't want to ask their granddaughter or grandson for help because they feel *paiseh* [embarassed]."} - P15, 19 y.o.). One older participant shared that Singaporeans might be afraid to lose face by asking for help and potentially avoid doing so as a result. (Researcher: \emph{“How can we teach older adults about the latest cyberthreats?"} Participant: \emph{“I think it is a Singaporean attitude, they think they know, but actually they don’t know… in a way the Chinese or Asians say, they want to save face.”}- P18, 75 y.o.). 
The common theme of "saving face" to avoid embarrassment when seeking help or learning could be a result of Singapore's highly collectivistic culture\cite{hofstede2020Culture}. This could be explained using behavioural models such as the Theory of Planned Behaviour\cite{ajzen1991theory} which posits social norms as a determinant for a person's behaviour.

\subsection{Privacy and Security Preferences}
\textbf{Contextual Preferences of Privacy}
The interviews highlighted that older adults were generally unaware about data flows and customizing of application permissions. However they expressed concerns about sharing personal data with application developers and wanted better control and visibility over data flows. They were generally comfortable with sharing of personal data with family members who could use the information to provide assistance and also expressed that their privacy preferences were influenced by that of their family members. (\textit{"For me, I am not working anymore, and I have nothing to hide...however, I try to mind my husband, my daughter, my son's interpretation of privacy"} - P7, 66 y.o.) and (\emph{“I respect their privacy and I’m careful not having my family in sight...”} - P19, 76 y.o.)  Although participants had customized preferences of data privacy ~\cite{camp2008beyond,hornung2017navigating}, it highlights the need to increase awareness about data, applications data might be collecting about them and the need to redesign privacy controls and data flow visualizations which are senior-friendly.

\textbf{Convenience vs Privacy \& Security}
Another theme which emerged among participants was convenience versus P \& S. Although adult participants were willing to forgo privacy for the sake of convenience at times, (\emph{“While I appreciate my privacy, I am still willing to compromise my privacy sometimes for the sake of convenience.”} P8, 44 y.o.), (\emph{“I think for connecting to the Internet, a certain level of privacy has to be given up...for the convenience of a connected world."} - P16, 23 y.o), they were generally concerned about older adults' convenient access to applications of all kinds, via the application store and advertisements-predominantly in Android phones-(\emph{“They may download apps on their phones for the sake of convenience or incentives that they promise.”} - P8, 44 y.o.). (\emph{“They install all the ... random shitty apps which pops up now and then.”} - P3, 34 y.o.). P15 who actively volunteers at the Digital Clinics\cite{IMDA2020} for older adults, found that many of them might \emph{" have random apps which are installed out of nowhere... some apps cause advertisements to popup and ... we help to remove them since they find it annoying.”}. Furthermore, almost all older adults participants exhibited little or no knowledge about how URL schemes worked and P18 suggested there should be more education about URLs (\emph{“ I think this is where education should come in ... how to check [interpret a URL] " } - P18, 75 y.o.)
With the ease of installing applications unknowingly and clicking on URLs without making informed decisions~\cite{Kelley2012}, an older adult's “blind spots”\cite{Frik2019} in knowledge about P \& S leaves them susceptible to installing malicious applications or giving away unintended personal data or falling for phishing scams(Figure~\ref{fig:summary}).
This brings about an opportunity for designers to increase  older adults' awareness about potential dangers and may be done by improving the inclusivity of existing safety mechanisms using frameworks such as the human-in-the-loop security framework\cite{Cranor2008}.
\section{Conclusion and Future Work}
In this paper, we identified challenges, including reliance of social circle, cultural and attitudinal factors which may hinder P \& S learning process and realized that although older adults had P \& S preferences while using their smartphones, opportunities arise for redesigning existing smartphone P \& S mechanisms to cater to the needs of older adults. We also identified contrasting perceptions of convenience vs P \& S between adults and older adults which could be explored as future work. As future works, we plan to formulate our findings and engage a multidisciplinary group of stakeholders to develop an inclusive technology probe~\cite{Hutchinson2003} to explore how we could allow older adults to make more informed P \& S decisions.  



\section{Acknowledgements}
The authors' gratitude goes to Dr. Hyowon Lee, Professor Jianying Zhou, Sujithra Raviselvam, Professor Kristin L. Wood and Nguyen Thi Ngoc for their inputs and guidance.
\bibliographystyle{plain}
\bibliography{usenix2020_SOUPS}

\end{document}